\documentclass[twocolumn,preprintnumbers,aps,prd]{revtex4}

\usepackage{amssymb,amstext,amsmath,bm,bbm}
\usepackage{xcolor}

\newcommand{\sout}[1]{}
\newcommand{\ssout}[1]{}
\newcommand{\red}[1]{#1}

\newcommand{\esout}[1]{}

\newcommand{\ket}[1]{{\vert{#1}\rangle}}

\newcommand{\realni}{\ensuremath{\mathbb{R}}}

\newcommand{\del}{\partial}

\newcommand{\lc}{\varepsilon}

\newcommand{\itGamma}{ {\varGamma} }                                    

\newcommand{\cD}{{\cal D}}

\begin{document}

\title{Equivalence Principle in Classical and Quantum Gravity}

\author{Nikola Paunkovi\'c$^{1}$}
 \email{npaunkov@math.tecnico.ulisboa.pt}

\author{Marko Vojinovi\'c$^{2}$}
 \email{vmarko@ipb.ac.rs}

\affiliation{
$^{1}$Instituto de Telecomuni\c{c}a\~oes and Departamento de Matem\'atica, Instituto Superior T\'ecnico, Universidade de Lisboa, Avenida Rovisco Pais 1, 1049-001, Lisboa, Portugal 
\\
$^{2}$Institute of Physics, University of Belgrade, Pregrevica 118, 11080 Belgrade, Serbia
}

\begin{abstract}
We give a general overview of various flavors of the equivalence principle in classical and quantum physics, with special emphasis on the so-called weak equivalence principle, and contrast its validity in mechanics versus field theory. We also discuss its generalisation to a theory of quantum gravity. \red{Our analysis suggests that only the strong equivalence principle can be considered fundamental enough to be generalised to a quantum gravity context, since all other flavors of equivalence principle hold only approximately already at the classical level.}
\end{abstract}



\maketitle

\section{Introduction}
\label{sec:introduction}

Quantum mechanics (QM) and general relativity (GR) are the two cornerstones of modern physics. Yet, merging them together in a quantum theory of gravity (QG) is still elusive despite nearly century long efforts of vast numbers of physicists and mathematicians. While the majority of the attempts were focused on trying to formulate the full theory of quantised gravity, such as  string theory, loop quantum gravity, non-commutative geometry, and causal set theory, to name a few, a number of recent studies embraced a rather more modest approach by exploring possible consequences of basic features and principles of QM and GR, and their status, in a tentative theory of QG. Acknowledging that the superposition principle, as a defining characteristic of any quantum theory, must be featured in QG as well, led to a number of papers studying gravity-matter entanglement~\cite{kay:98,oni:wan:16,bru:16,bos:etal:17,mar:ved:17,mar:ved:18,pau:voj:18}, genuine indefinite causal orders~\cite{ore:cos:bru:12,ara:bra:cos:fei:gia:bru:15,vil:17,ore:19,pau:voj:20,vil:col:22,vil:ren:22,orm:van:bar:22}, quantum reference frames~\cite{gia:cas:bru:19,van:hoh:gia:cas:20,kru:hoh:mul:21,ahm:etal:22,ham:kab:cas:bru:21} \red{and deformations of Lorentz symmetry \cite{col:kos:98,kos:rus:11,ame:12,ame:pal:ron:ami:20,tor:ant:mir:19}}, to name a few major research directions. Exploring spatial superpositions of masses, and in general of gravitational fields, led to the analysis of the status of various versions of the equivalence principle, and their exact formulations in the context of QG. In particular, in~\cite{pip:pau:voj:19} it was shown that the weak equivalence principle (WEP) should generically be violated in the presence of a specific type of superpositions of gravitational fields, \red{describing small quantum fluctuations around a dominant classical geometry. On the other hand, a number of recent studies propose generalisations of WEP to QG framework (see for example~\cite{gia:cas:bru:19,gia:bru:20,ham:kab:cas:bru:21,gia:bru:22,mar:ved:20,mar:ved:21,mar:ved:22}), arguing that it remains satisfied in such scenarios, a result {\em seemingly} at odds with~\cite{pip:pau:voj:19}} (for details, see the discussion from Section~\ref{sec:conclusions}).

Modern formulation of WEP is given in terms of a {\em test particle} and its {\em trajectory}: it is a {\em theorem} within the mathematical formulation of GR stating that the trajectory of a test particle satisfies the so-called geodesic equation~\cite{EinsteinInfeldHoffmann,Mathisson,Papapetrou,Tulczyjew,Taub,Dixon1,Dixon2,Dixon3,Dixon4,Dixon5,YasskinStoeger,ShirafujiNomuraHayashiOne,ShirafujiNomuraHayashiTwo,VasilicVojinovicJHEP,VasilicVojinovicCetiri}, while it is {\em violated} within the context of QG, as shown in~\cite{pip:pau:voj:19}. In this paper, we present a brief overview of WEP in GR and a critical analysis of the notions of particle and trajectory in both classical and quantum mechanics, as well as in the corresponding field theories.\ssout{, with the aim of generalising and studying WEP in the context of QG} \red{Our analysis demostrates that WEP, as well as all other flavors of the equivalence principle (EP) aside from the strong one (SEP), hold only approximately. From this we conclude that neither WEP nor any other flavor of EP (aside from SEP) can be considered a viable candidate for generalisation to the quantum gravity framework.}

The paper is organised as follows. In Section~\ref{sec:wep}, we give a brief historical overview of various flavours of the equivalence principle, with a focus on WEP. In Section~\ref{sec:trajectory_in_mechanics} we analyse the notion of a trajectory in classical and quantum mechanics, while in Section~\ref{sec:particle_in_gr} we discuss the notion of a particle in field theory and QG. Finally, in Conclusions, we briefly review and discuss our results, and present possible future lines of research.

\section{Equivalence principle in general relativity}
\label{sec:wep}

The equivalence principle is one of the most fundamental principles in modern physics. It is one of the two cornerstone building blocks for GR, the other being the principle of general relativity. While its importance is well understood in the context of gravity, it is often underappreciated in the context of other fundamental interactions. In addition, there have been numerous studies and everlasting debates whether EP holds also in quantum physics, if it should be generalised to include quantum phenomena or not, etc. Finally, EP has been historically formulated in a vast number of different ways, which are often not mutually equivalent, leading to a lot of confusion about the actual statement of the principle and its physical content \cite{AcciolyPaszko,Longhi,Chowdhury,Rosi,zyc:bru:17,ana:hu:18,har:20}. Given the importance of EP, and the amount of confusion around it, it is important to try and help clarify these issues.

The equivalence principle is best introduced by stating its purpose --- in its traditional sense, the purpose of EP is to {\em prescribe the interaction between gravity and all other fields in nature, collectively called matter} \red{(by ``matter'' we assume not just fermionic and scalar fields, but also gauge vector bosons, i.e., nongravitational interaction fields)}. This is important to state explicitly, since EP is often mistakenly portrayed as a property of gravity alone, without any reference to matter. In a more general, less traditional, and often not appreciated sense, the purpose of EP is to prescribe the interaction between {\em any gauge field} and all other fields in nature \red{(namely fermionic and scalar matter, as well as other gauge fields, including gravity)}, which we will reflect on briefly in the case of electrodynamics below.

Given such a purpose, let us for the moment concentrate on the gravitational version of EP, and provide its modern formulation, as it is known and understood today. The statement of the equivalence principle is the following:
\begin{quote}
\medskip
  {\em The equations of motion for matter coupled to gravity remain locally identical to the equations of motion for matter in the absence of gravity.}
\medskip
\end{quote}
This kind of statement requires some unpacking and comments.
\begin{itemize}
\item When comparing the equations of motion in the presence and in the absence of gravity, the claim that they remain identical may naively suggest that gravity does not influence the motion of matter in any way whatsoever. But on closer inspection, the statement is that the two sets of equations remain {\em locally} identical, emphasizing that the notion of locality is a crucial feature of the EP. While equations of motion are already local in nature (since they are usually expressed as partial differential equations of finite order), the actual interaction between matter and gravity enters only when {\em integrating} those equations to find a solution \red{(see Appendix \ref{sec:app} for a detailed example)}. \ssout{In other words, the statement of EP is that interaction between matter and gravity appears in a specifically nonlocal sense.}
 
\item In order to compare the equations of motion for matter in the presence of gravity to those in its absence, the equations themselves need to be put in a suitable form (typically expressed in general curvilinear coordinates, as tensor equations). The statement of EP relies on a theorem that this can always be done, first noted by Erich Kretschmann \cite{kretschmann}. 
  
\item Despite being dominantly a statement about the interaction between matter and gravity, EP also implicitly suggests that the best way to describe the gravitational field is as a property of geometry of spacetime, such as its metric \cite{mis:tho:whe:73}. In that setup, EP can be reformulated as a statement of {\em minimal coupling} between gravity and matter, stating that equations of motion for matter may depend on the spacetime metric and its first derivatives, but not on its (antisymmetrised) second derivatives, i.e., the {\em spacetime curvature does not explicitly appear in the equations of motion for matter}. 
  
\item The generalisation of EP to other gauge fields is completely straightforward, by replacing the role of gravity with some other gauge field, and suitably redefining what matter is. For example, in electrodynamics, the EP can be formulated as follows:
\begin{quote}
\medskip
  {\em The equations of motion for matter coupled to the electromagnetic field remain locally identical to the equations of motion for matter in the absence of the electromagnetic field.}
\medskip
\end{quote}
This statement can also be suitably reformulated as the minimal coupling between the electromagnetic (EM) field and matter, i.e., coupling matter to the electromagnetic potential $A_\mu$ but not to the corresponding field strength $F_{\mu\nu} = \del_\mu A_\nu - \del_\nu A_\mu$. This is in fact the standard way EM field is coupled to matter \red{(see Appendix \ref{sec:app} for an illustrative example)}. Even more generally, the gauge field sector of the whole Standard Model of elementary particles (SM) is built using the minimal coupling prescription, meaning that the suitably generalised version of the EP actually prescribes the interaction between matter and all fundamental interactions in nature, namely strong, weak, electromagnetic and gravitational. In this sense, EP is a cornerstone principle for the whole fundamental physics, as we understand it today. 
\end{itemize}
Of course, much more can be said about the statement of EP, its consequences, and various other details. However, in this work our attention will focus on the so-called {\em weak equivalence principle} (WEP), which is a reformulation of EP applied to matter which consists of mechanical particles. To that end, it is important to understand various flavors and reformulations of EP that have appeared through history.

As any deep concept in physics, EP has been expressed historically through painstaking cycle of formulating it in a precise way, understanding the formulation, understanding the drawbacks of that formulation, coming up with a better formulation, and repeating. In this sense, EP, as quoted above, is a modern product of long and meticulous refinement over several generations of scientists. Needless to say, each step in that process made its way into contemporary physics textbooks, leading to a plethora of different formulations of EP that have accummulated in the literature over the years. This can bring about a lot of confusion about what EP actually states \cite{AcciolyPaszko,Longhi,Chowdhury,Rosi}, since various formulations from old and new literature may often be not merely phrased differently, but in fact substantively inequivalent. To that end, let us comment on several most common historical statements of EP (for a more detailed historical overview and classification, see \cite{oco:cal:11,cas:lib:son:15}), and their relationship with the modern version:
\begin{itemize}
\item {\em Equality of gravitational and inertial mass.} This is one of the oldest variants of EP, going back to Newton's law of universal gravitation. The statement claims that the ``gravitational charge'' of a body is the same as body's resistance to acceleration, in the sense that the mass appearing on the left-hand side of Newton's second law of motion exactly cancels the mass appearing in the Newton's gravitational force law on the right-hand side. This allows one to relate it to the modern version of EP, in the sense that a suitably accelerated observer could rewrite the Newton's law of motion as the equation for a free particle, exploiting the cancellation of the ``intertial force'' and the gravitational force on the right-hand side of the equation. Unfortunately, this version of EP is intrinsically nonrelativistic, and applicable only in the context of Newtonian gravity, since already in GR the source of gravity becomes the full stress-energy tensor of matter fields, rather than just the total mass. Finally, this principle obviously fails when applied to photons, as demonstrated by the gravitational bending of light. 
  
\item {\em Universality of free fall.} Going back all the way to Galileo, this statement claims that the interaction between matter and gravity does not depend on any intrinsic property of matter itself, such as its mass, angular momentum, chemical composition, temperature, or any other property, leading to the idea that gravity couples universally (i.e., in the same way) to all matter. Formulated from experimental observations by Galileo, its validity is related to the quality of experiments used to verify it. As we shall see below, in a precise enough setting, one can experimentally observe direct coupling between the angular momentum of a body and spacetime curvature \cite{EinsteinInfeldHoffmann,Mathisson,Papapetrou,Tulczyjew,Taub,Dixon1,Dixon2,Dixon3,Dixon4,Dixon5,YasskinStoeger,ShirafujiNomuraHayashiOne,ShirafujiNomuraHayashiTwo,VasilicVojinovicJHEP,VasilicVojinovicCetiri}, invalidating the statement. 
  
\item {\em Local equality between gravity and inertia.} Often called Einstein's equivalence principle, the statement claims that a local and suitably isolated observer cannot distinguish between accelerating and being at rest in a uniform gravitational field. While this statement is much closer in spirit to the modern formulation of EP, it obscures the crucial aspect of the principle --- coupling of matter to gravity. Namely, in this formulation it is merely implicit that the only way an observer can {\em attempt to distinguish} gravity from inertia is by making local experiments using some form of {\em matter}, i.e., studying the equations of motion of matter in the two situations and trying to distinguish them by observing whether or not matter behaves differently. Moreover, the statement is often discussed in the context of mechanics, arguing that any given particle does not distinguish between gravity and inertia. This has two main pitfalls --- first, the reliance on particles is very misleading (as we will discuss below in much more detail), and second, it implicitly suggests that gravity and inertia are the same phenomenon, which is completely false. Namely, inertia can be understood as a specific form of gravity, but a general gravitational field cannot be simulated by inertia, since inertia cannot account for tidal effects of inhomogeneous configurations of gravity. 

\item {\em Weak equivalence principle.} Stating that the equations of motion of particles do not depend on spacetime curvature, or equivalently, that the motion of a free particle is always a geodesic trajectory in spacetime, WEP is in fact an application of modern EP to mechanical point-like particles (i.e., test particles). One can argue that, as far as the notion of a point-like particle is a well defined concept in physics, WEP is a good principle. Nevertheless, as we will discuss below in detail, the notion of a point-like particle is an idealisation that does not actually have any counterpart in reality, in either classical or quantum physics. Regarding a realistic particle (with nonzero size), WEP {\em never holds}, due to the explicit effect of gravitational tidal forces across the particle's size. In this sense, WEP can be considered at best an {\em approximate} principle, which can be assumed to hold only in situations where particle size can be approximated to zero. 
  
\item {\em Strong equivalence principle.} This version of the principle states that the equations of motion of all fundamental fields in nature do not depend on spacetime curvature (see~\cite{mis:tho:whe:73}, Section 16.2, page 387). To the best of our knowledge so far, fields are indeed the most fundamental building blocks in modern physics (such as SM), while the strength of the gravitational field is indeed described by spacetime curvature (as in GR). In this sense, the statement of SEP is actually an instance of EP applied to field theory, and as such equivalent to the modern statement of EP. So far, all our knowledge of natural phenomena is consistent with the validity of SEP. 
\end{itemize}
As can be seen from the above review, various formulations of EP are both mutually inequivalent and have different domains of applicability. \red{Specifically, only SEP holds universally, while all other flavors of EP hold only approximately.} In the remainder of the paper, we focus on the study of WEP, since recently it gained a lot of attention in the literature~\cite{gia:bru:20,ham:kab:cas:bru:21,gia:bru:22,mar:ved:20,mar:ved:22}, primarily in the context of its generalisation to a ``quantum WEP'', and in the context of a related question of particle motion in a quantum superposition of different gravitational configurations, the latter being a scenario that naturally arises in QG. Since WEP is stated in terms of a test particle and its trajectory, in order to try and generalise it to the scope of QG one should first analyse these two notions in classical and quantum mechanics and field theory in more detail.

\section{The notion of trajectory in classical and quantum mechanics}
\label{sec:trajectory_in_mechanics}

A trajectory of a physical system in three-dimensional space is a set of points that form a line, usually parametrised by time. More formally, a trajectory is a set $\{(x(t),y(t),z(t)) \in \mathbb{R}^3 | t \in [t_i, t_f] \subset \mathbb R \land t_i < t_f \}$, given by three smooth functions $x, y, z : \mathbb R \mapsto \mathbb R$. Depending on the nature of the system, the choice of points that form its trajectory may vary.

In classical mechanics, one often considers an ideal ``point-like particle'' localised in one spatial point $(x(t),y(t),z(t))$ at each moment of time $t$, in which case the choice of the points forming the system's trajectory is obvious. In case of systems continuously spread over certain volume (``rigid bodies'', or ``objects''), or composite systems consisting of several point-like particles or bodies, it is natural to consider their centres of mass as points that form the trajectory. While this definition is natural, widely accepted, and formally applicable to any classical mechanical system, there are cases in which the very notion of a trajectory looses its intuitive, as well as useful, meaning.

Consider for example an electrical dipole, i.e., a system of two point-like particles with equal masses and opposite electrical charges, separated by the distance $\ell(t)$. As long as this distance stays ``small'' and does not vary significantly with time, the notion of a trajectory of a dipole, defined as the set of centres of mass of the two particles, does meet our intuition, and can be useful. Informally, if the trajectories of each of the two particles are ``close'' to each other, they can be approximated, and consequently represented, by the trajectory of the system's centre of mass. But if the separate trajectories of the two particles diverge, one going to the ``left'', and another to the ``right'', one could hardy talk of a trajectory of such a composite system, although the set of locations of its centres of mass is still well defined. \red{In fact, the dipole itself ceases to make physical sense when the distance between its constituents is large.}

Moving to the realm of quantum mechanics, due to the superposition principle, even the ideal point-like particles do not have a well defined position, which is further quantified by the famous Heisenberg uncertainty relations. Thus, the trajectory of point-like particles (and any system that in a given regime can be approximated to be point-like) is defined as a set of expectation values of the position operator. Like in the case of composite classical systems, here as well the definition of a trajectory of a point-like particle is mathematically always well defined, yet for the very similar reason is applicable only to certain cases. Namely, in order to give a useful meaning to the above definition of trajectory, the system considered must be {\em well localised}. Consider for example the double-slit experiment, in which the point-like particle is highly delocalised, so that we say that {\em its trajectory is not well defined}, even though the set of the expectation values of the position operator is.

We see that, while in mechanics both the notions of a particle and its trajectory are rather straightforward and always well defined, the latter make sense only if our system is well localised in space \red{(see for example \cite{vio:ono:97}, where the authors analyse the effects of wave-packet spreading to the notion of a trajectory)}.

\section{The notion of a particle in field theory}
\label{sec:particle_in_gr}

While in classical mechanics a point-like particle is always well localised, we have seen that in the quantum case one must introduce an additional constraint in order for it to be considered localised --- the particle should be represented by a wave-packet. The source for this requirement lies in the fact that quantum particles, although mechanical, are represented by a {\em wavefunction}. Thus, it is only to be expected that when moving to the realm of the field ontology, the notion of a particle becomes even more involved and technical.

In field theory, the fundamental concept is the {\em field}, rather than a particle. The notion of a particle is considered a derived concept, and in fact in QFT one can distinguish two vastly different phenomena that are called ``particles''.

The first notion of a particle is an elementary excitation of a free field. For example, the state
\begin{equation*}
\ket{\Psi} = \hat{a}^\dag(\vec{k}) \ket{0} \,,
\end{equation*}
is called a {\em single particle state} of the field, or a {\em plane-wave particle}. It has the following properties:
\begin{itemize}
\item It is an eigenstate of the {\em particle number operator} for the eigenvalue $1$.
\item It has a sharp value of the momentum $\vec{k}$, and corresponds to a completely delocalised plane wave configuration of the field.
\item It has no centre of mass, and no concept of ``position'' in space, since the ``position operator'' is not a well defined concept for the field.
\item States of this kind are said to describe {\em elementary particles}\red{, understood as asymptotic free states of past and future infinity, in the context of the $S$-matrix for scattering processes.} \ssout{For a real scalar field, an example}\red{An example of a real scalar particle of this type} would be the {\em Higgs particle}. For fields of other types (Dirac fields, vector fields, etc.) examples would be an {\em electron}, a {\em photon}, a {\em neutrino}, an asymptotically free {\em quark}, and so on. Essentially, all particles tabulated in the Standard Model of elementary particles are of this type.
\end{itemize}
\red{Note that all the above notions are defined within the scope of free field theory, and do not carry over to interacting field theory. In other words, free field theory is a convenient idealisation, which does not really reflect realistic physics. One should therefore understand the concept of plane-wave particle in this sense, merely as a convenient mathematical approximation. Moreover, the particle number operator is not an invariant quantity, as demonstrated by the Unruh effect. We should also emphasise that in an interacting QFT, the proper way to understand the notion of a particle is as a localised wave-packet, interacting with its virtual particle cloud, which does have a position in space and whose momentum is defined through its group velocity. In this sense, the particle as a wave-packet could be better interpreted as a kink, discussed below.}

The second notion of the particle in field theory is a bound state of fields, also called a {\em kink solution}. This requires an interacting theory, since interactions are necessary to form bound states. This kind of configuration of fields has the following properties:
\begin{itemize}
\item It is not an eigenstate of the particle number operator, and the expectation value of this operator is typically different from $1$.
\item It is usually well localised in space, and does not have a sharp value of momentum.
\item As long as the kink maintains a stable configuration (i.e., as long as it does not decay), one can in principle assign to it the concept of {\em size}, and as a consequence also the concepts of {\em centre of mass}, {\em position in space}, and {\em trajectory}. In this sense, a kink can play the role of a test particle.
\item States of this kind are said to describe {\em composite particles}. Given an interacting theory such as the Standard Model, under certain circumstances quarks and gluons form bound states called a {\em proton} and a {\em neutron}. Moreover, protons and neutrons further form bound states called {\em atomic nuclei}, which together with electrons and photons form {\em atoms}, {\em molecules}, and so on.
\end{itemize}
For a kink, the notions of centre of mass, position in space and size are described only as classical concepts, i.e., as expectation values of certain field operators, such as the stress-energy tensor. Moreover, given the nonzero size of the kink, its centre of mass and position are not uniquely defined, even classically, since in relativity different observers would assign different points as the centre of mass.

Given the two notions of particles in QFT, one can describe two different corresponding notions of WEP. In principle, one first needs to apply SEP in order to couple the matter fields to gravity, at the fundamental level. Assuming this is done, the motions of both the plane-wave particles and kink particles can be derived from the combined set of Einstein's equations and matter field equations, without any appeal to any notion of WEP. In this sense, once the trajectory of the particle in the background gravitational field has been determined from the field equations, one can verify {\em as a theorem} whether the particle satisfies WEP or not.

Specifically, in the case of a matter field coupled to general relativity such that it  locally resembles a plane wave, one can apply the WKB approximation to demonstrate that the wave $4$-vector $k^{\mu}(x)$, orthogonal to the wavefront at its every point $x\in \realni^4$, will satisfy a geodesic equation,
\begin{equation} \label{eq:geodesic_plane_wave}
k^\mu(x) \nabla_\mu k^\lambda(x) = 0\,.
\end{equation}
However, given that the plane-wave particle is completely delocalised in space, the fact that the wave $4$-vector satisfies the geodesic equation could hardly be interpreted as ``the particle following a geodesic trajectory'', and thus obeying WEP. Indeed, identifying the vector field orthogonal to the wavefront to the notion of ``particle's trajectory'' is at best an abuse of terminology.

Next, in the case of the kink particle coupled to general relativity, one assumes the configuration of the background gravitational field is such that the particle maintains its structure and that its size can be completely neglected. One can then apply the procedure given in \cite{EinsteinInfeldHoffmann,Mathisson,Papapetrou,Tulczyjew,Taub,Dixon1,Dixon2,Dixon3,Dixon4,Dixon5,YasskinStoeger,ShirafujiNomuraHayashiOne,ShirafujiNomuraHayashiTwo,VasilicVojinovicJHEP,VasilicVojinovicCetiri,pip:pau:voj:19} to demonstrate that the $4$-vector $u^\mu(\tau)$, tangent to the kink's world line (i.e., its trajectory), will satisfy a geodesic equation ($\tau \in \realni $ represents kink's proper time),
\begin{equation} \label{eq:geodesic_kink}
u^\mu(\tau) \nabla_\mu u^\lambda(\tau) = 0\,.
\end{equation}
Thus, one concludes that the kink obeys WEP as a {\em theorem} in field theory, without the necessity to actually postulate it.

Note the crucial difference between equations (\ref{eq:geodesic_plane_wave}) and (\ref{eq:geodesic_kink}) --- while the former features $4$-dimensional variable $x$, the latter is given in terms of only $1$-dimensional proper time $\tau$. This reflects the fact that the plane-wave particle is a highly delocalised object, with no well defined position and trajectory, while the kink is a highly localised object, with well defined position and trajectory. As a consequence, WEP can be formulated only for the kink, and not for the plane-wave particle.

In the case of the kink it is also important to emphasise that the zero size approximation of the kink is crucial. Namely, without this assumption, the particle will feel the tidal forces of gravity across its size, effectively coupling its angular momentum $J^{\mu\nu}(\tau)$ to the curvature of the background gravitational field \cite{EinsteinInfeldHoffmann,Mathisson,Papapetrou,Tulczyjew,Taub,Dixon1,Dixon2,Dixon3,Dixon4,Dixon5,YasskinStoeger,ShirafujiNomuraHayashiOne,ShirafujiNomuraHayashiTwo,VasilicVojinovicJHEP,VasilicVojinovicCetiri} \red{(see also \cite{ply:98} for a more refined analysis of tidal effects)}. This will give rise to an equation of motion for the kink of the form
\begin{equation} \label{eq:PapapetrouEq}
u^\mu(\tau) \nabla_\mu u^\lambda(\tau) = R^\lambda{}_{\mu\rho\sigma} u^\mu(\tau) J^{\rho\sigma}(\tau) \,,
\end{equation}
which features explicit coupling to curvature (absent from (\ref{eq:geodesic_kink})) and thus fails to obey WEP. In this sense, for realistic kink solutions WEP is {\em always violated}, and can be considered to hold only as an approximation, when the size of the particle can be completely neglected compared to the radius of curvature of the background gravitational field.  If in addition the kink has negligible total energy, it can be used as a point-like test particle.

In the above discussion, while matter fields are described as quantum, using QFT, the background gravitational field is considered to be completely classical. It should therefore not be surprising that WEP may fail to hold if one allows the gravitatational field to be quantum, like matter fields, and one needs to revisit all steps of the above analysis from the perspective of QG. In fact, the case of the kink particle has been studied in precisely this scenario \cite{pip:pau:voj:19}, and it has been shown that if the background gravitational field is in a specific type of quantum superposition of different configurations, the kink will fail to obey WEP even in the zero size approximation. Simply put, the equation of motion for the kink will contain extra terms due to the interference effects between superposed configurations of gravity, giving rise to an effective force that pushes the kink off the geodesic trajectory. And of course, similar to the case of the classical gravity, the resulting conclusion is a {\em theorem}, which follows from the fundamental field equations of the theory. \red{One of the assumptions of that theorem is that the field equations allow for kink solutions in the first place. Namely, it is entirely possible that in quantum gravity particles cannot be localised at all, as opposed to the classical case where such an approximation can be feasible. If that is the case, then one cannot even formulate (i.e., generalise from classical theory) the notion of WEP in a quantum gravity setup. However, one can instead assume that kink solutions do exist, as was done in \cite{pip:pau:voj:19}, where a particular superposition of gravitational fields was considered, describing small quantum fluctuations around a dominant classical geometry. It was then argued that such superpositions are compatible with the approximation of a well-defined localised particle (see the discussion around equations (2.2) and (3.15), as well as Section 3.4 of that paper). As it turns out, even in such cases the trajectory of the kink fails to obey WEP. Therefore, the generalisations of WEP and other approximate versions of EP are not the best candidates for analysing the properties of quantum gravity.

Moreover, the assumption of a well-defined notion of a particle in the QG framework can also be supported from the point of view of nonrelativistic limit. Namely, in \cite{bos:etal:17,mar:ved:17} an experiment was proposed in which the effects of QG fluctuations are expected to be observable, by measuring the motion of nonrelativistic particles. Furthermore, an extension of this experiment was also suggested \cite{bos:maz:sch:tor:22}, which aims to determine the potential difference between gravitational and inertial masses of those particles in such a setup. In fact, the relation between the two types of masses in the nonrelativistic limit has also been previously analysed in \cite{pip:pau:voj:19}, predicting their difference due to quantum fluctuations of geometry. In this sense, the notion of a kink should make sense even in the QG setup, at least in the nonrelativistic limit.
}

For the case of the plane-wave particle travelling through the superposed background of two gravitational field configurations, the analysis \red{of the equation of motion for the wave-vector field $k^\mu(x)$, in the sense of \cite{EinsteinInfeldHoffmann,Mathisson,Papapetrou,Tulczyjew,Taub,Dixon1,Dixon2,Dixon3,Dixon4,Dixon5,YasskinStoeger,ShirafujiNomuraHayashiOne,ShirafujiNomuraHayashiTwo,VasilicVojinovicJHEP,VasilicVojinovicCetiri,pip:pau:voj:19},} has not been done so far (to the best of our knowledge). But in principle one can expect a similar interference term to appear in the WKB analysis, and give rise to a non-geodesic equation for the wave $4$-vector as well. In this sense, it is to be expected that generically even the wavefronts of such plane-wave particles would fail to obey WEP.

\section{Conclusions and discussion}
\label{sec:conclusions}

In this paper, we give an overview of the equivalence principle and its various flavors formulated historically, with a special emphasis on the weak equivalence principle. \ssout{In order to generalise WEP to QG,} We performed a critical analysis of the notions of particle and trajectory in various frameworks of physics, showing that the notion of point-like particle and its trajectory are not always well defined. This in turn suggests that WEP\ssout{ is not a good} \red{might not be the best} starting point for generalisation to QG, as we argue in more detail below.

As discussed in Section~\ref{sec:particle_in_gr}, in~\cite{pip:pau:voj:19} it was shown that if superpositions of states of gravity and matter are allowed, WEP can be violated. It is important to note that the cases considered in~\cite{pip:pau:voj:19} feature a specific type of superpositions of three groups of states: the first consists of a single so-called dominant state --- a classical state whose expectation values of the metric and the stress-energy tensors satisfy Einstein field equations; the second consists of states similar to the dominant one, with arbitrary coefficients; and the third consists of states quasi-orthogonal to the dominant one, but with negligible coefficients. Only then one\ssout{ can} \red{may} talk of a (well localised) trajectory of the test particle in the overall superposed state and consequently about the straightforward generalisation of the classical WEP to the realm of QG. Considering that for the dominant state, being classical, the trajectory of the test particle follows the corresponding geodesic, we see that in the superposed state its trajectory would {\em deviate from the geodesic of the dominant state}, thus violating WEP. Note that, as discussed in Section~\ref{sec:particle_in_gr}, this deviation, in addition to classically weighted trajectories of the individual branches, also features purely quantum (i.e., off-diagonal) interference terms.

\ssout{On the other hand, in recent studies}\esout{~\cite{gia:bru:20,ham:kab:cas:bru:21,gia:bru:22,mar:ved:20,mar:ved:22} }\ssout{it was proposed that WEP is {\em not violated} in QG, thus being at odds with our result. The authors of the mentioned studies considered superpositions of an arbitrary number of classical quasi-orthogonal states with arbitrary coefficients. Their argument goes as follows:}
\red{On the other hand, a number of recent studies propose generalisations of WEP to QG framework, arguing that it remains satisfied in such scenarios, a result {\em seemingly} at odds with~\cite{pip:pau:voj:19}. For example, in~\cite{mar:ved:20,mar:ved:21,mar:ved:22} the authors consider superpositions of an arbitrary number of classical quasi-orthogonal states with arbitrary coefficients, arguing that} since WEP is valid in each classical branch, it is valid in their superposition as well. If taken as a {\em definition} of what does it mean that a certain principle is satisfied in a superposition of different quantum states, then the above statement is manifestly true. But, as such, being a definition, it tells little about physics --- it merely rephrases one statement (``principle A is separately satisfied in all branches of a superposition'') with another, simpler (``principle A is satisfied in a superposition''). \red{Namely, note that in \cite{mar:ved:20,mar:ved:21}, such generalised version of EP plays no functional role in the analyses done in those papers. What does play a functional role is the statement of one version of classical EP (specifically, local equality between gravity and inertia) applied to each particular state in a superposition. All physically relevant (and otherwise interesting) conclusions of the two papers could be equally obtained without ever talking about the generalised EP. In addition, in \cite{mar:ved:22} EP itself is not even the main focus of the paper, and its generalisation is just introduced in analogy to the analysis of the conservation laws, which is itself an interesting topic.
On the other hand, in the case of weakly superposed gravitational fields, such as in proposed experiments \cite{mar:ved:17,bos:etal:17}, the violation of the equality of inertial and gravitational masses is to be expected \cite{pip:pau:voj:19,bos:maz:sch:tor:22}. Moreover, following the spirit of the above definition, one could be misled to conclude that the notions of particle's position and trajectory are always well defined, as long as they are defined in each (quasi-classical) branch of the superposition.
}

\red{An alternative approach to the generalisation of EP to quantum domain was proposed in \cite{gia:cas:bru:19,gia:bru:20,ham:kab:cas:bru:21,gia:bru:22}. In those works, the authors discuss the coupling of a spatially delocalised wave-particle to gravity, with the aim of generalising such a scenario to QG. To that end, they prove a theorem which essentially states that for such a delocalised wave-particle, even when it is entangled with the gravitational field, one can always find a quantum reference frame  transformation, such that in the vicinity of a given spacetime point one has a locally inertial coordinate system. The theorem employs the novel techniques of quantum reference frames (QRF) to generalise to quantum domain the well-known result from differential geometry, that in the infinitesimal neighbourhood of any spacetime point one can always choose a locally inertial coordinate system.

 The authors then employ the theorem to generalise one flavour of EP to quantum domain. Specifically, even if the wave-particle is entangled with the gravitational field, one can use the appropriate QRF transformation to switch to a locally inertial coordinate system, and then in that system ``all the (nongravitational) laws of physics must take on their familiar non-relativistic form''. Here, to the best of our understanding, the phrase ``nongravitational laws of physics'' refers to the equations of motion for a quantum-mechanical wave-particle, while ``non-relativistic form'' means that these equations of motion take the same form as in special-relativistic context.

Our understanding is that the above wave-particle generalisation of EP lies somewhere ``in between'' mechanics and field theory, i.e., it is in a sense stronger than WEP, which discusses particles, but weaker than SEP, which discusses full-blown matter fields. Since it refers to wave-particles rather than kinks, our analysis of WEP and its reliance on the particle trajectory does not apply to this version of EP.
 
The methodology in \cite{gia:cas:bru:19,gia:bru:20,ham:kab:cas:bru:21,gia:bru:22} is that one should try to generalise even approximate flavours of EP, as a stopgap result in a bigger research programme, in the hope that they may still shed some light on QG. This is of course a legitimate methodology, and from that point of view these kinds of generalisations of EP to quantum domain represent interesting results. Nevertheless, we also believe it would be preferable to formulate a generalisation of SEP, and in a way which does not appeal to reference frames at all, since that would be closer to the essence of the statement of EP, as discussed \mbox{in~Section~\ref{sec:wep}}.
}

\ssout{Finally, we would like to argue that the very notion of WEP is ill-defined in such ``extreme'' scenarios of superposed quasi-orthogonal classical states, and should not be even discussed. After all, we have seen that what we call a principle is nothing but a theorem in GR about the approximate trajectory of a point-like test particle, both of which are not always possible to define. Consequently, studying generalisations of WEP might not be very useful for inferring the features of QG. In our opinion, it would be better to instead study other principles and their possible generalisations to QG, such as SEP (see Section 4.2 in \cite{pip:pau:voj:19}), background independence, quantum nonlocality, and so on.}

\red{To conclude, our analysis suggests that, instead of trying to generalise various approximate formulations of EP, one should rather talk of operationally verifiable statements regarding the (in)equality of gravitational and inertial masses, possible deviation from the geodesic motion of test particles, universality of free fall, etc., and study other principles and their possible generalisations to QG, such as SEP (see Section 4.2 in \cite{pip:pau:voj:19}), background independence, quantum nonlocality, and so on.}

\vspace*{-0.3cm}

\begin{acknowledgments}
The authors wish to thank \v{C}aslav Brukner, Flaminia Giacomini, Chiara Marletto and Vlatko Vedral for useful discussions. MV is also indebted to Milovan Vasili\'c and Igor Salom for clarifications regarding the notion of symmetry localisation.

NP's work was partially supported by SQIG -- Security and Quantum Information Group of Instituto de Telecomunica\c{c}\~oes, by Programme (COMPETE 2020) of the Portugal 2020 framework [Project Q.DOT with Nr.\ 039728 (POCI-01-0247-FEDER-039728)] and the Funda\c{c}\~ao para a Ci\^{e}ncia e a Tecnologia (FCT) through national funds, by FEDER, COMPETE 2020, and by Regional Operational Program of Lisbon, under UIDB/50008/2020 (actions QuRUNNER, QUESTS), Projects QuantumMining POCI-01-0145-FEDER-031826, PRE\-DICT PTDC/CCI-CIF/29877/2017, CERN/FIS-PAR/0023/2019, QuantumPrime PTDC/EEI-TEL/8017/2020, as well as the FCT Est\'{i}mulo ao Emprego Cient\'{i}fico grant no. CEECIND/04594/2017/CP1393/CT000. 

MV was supported by the Ministry of Education, Science and Technological Development of the Republic of Serbia, and by the Science Fund of the Republic of Serbia, grant 7745968, ``Quantum Gravity from Higher Gauge Theory 2021'' --- QGHG-2021. The contents of this publication are the sole responsibility of the authors and can in no way be taken to reflect the views of the Science Fund of the Republic of Serbia.
\end{acknowledgments}

\onecolumngrid

\appendix

\red{

\section{}\label{sec:app}

Here we provide a detailed example of the two applications of the EP. First we discuss the gravitational EP and apply it to a real scalar field, giving all mathematical details and discussing various related aspects such as locality, symmetry localisation, and so on. Then we turn to the application of the gauge field generalisation of EP, using electrodynamics as an example. We describe how one can couple matter to an EM field, mimicking the previous gravitational example, and emphasize the analogy between the gravitational and EM case at each step. Note also that the non-Abelian gauge fields can be studied in exactly the same way. Finally, we discuss the case of test particles, and the violation of the WEP in both gravitational and electromagnetic cases.

Throughout this section, we assume that the Minkowski metric $\eta_{\mu\nu}$ has signature $(-,+,+,+)$.

\subsection{The gravitational case}

Let us begin with an example of a real scalar field in Minkowski spacetime, and apply the equivalence principle by coupling it to gravity. The equation of motion in this case is the ordinary Klein-Gordon equation,
\begin{equation} \label{eq:flatKG}
\left( \eta^{\mu\nu} \del_{\mu}\del_{\nu} - m^2 \right) \phi(x) = 0\,.	
\end{equation}
As it stands, it describes the free scalar field in Minkowski spacetime, in an inertial coordinate system. In order to couple it to gravity (in the framework of GR), we first rewrite this equation into an arbitrary curvilinear coordinate system, as
\begin{equation} \label{eq:curvilinearKG}
\left( \tilde{g}^{\mu\nu} \tilde{\nabla}_{\mu}\tilde{\nabla}_{\nu} - m^2 \right) \phi(\tilde{x}) = 0\,.	
\end{equation}
Here the covariant derivative $\tilde{\nabla}_{\mu}$ is defined in terms of the Levi-Civita connection,
\begin{equation} \label{eq:curvilinearConnection}
\tilde{\itGamma}^\lambda{}_{\mu\nu} = \frac{1}{2} \tilde{g}^{\lambda\sigma} \left( \del_\mu \tilde{g}_{\nu\sigma} + \del_\nu \tilde{g}_{\mu\sigma} - \del_\sigma \tilde{g}_{\mu\nu} \right)\,,
\end{equation}
which is in turn defined in terms of the curvilinear Minkowski metric $\tilde{g}_{\mu\nu}$. Note that the tilde symbol denotes the fact that this metric has been obtained by a coordinate transformation $\tilde{x}^\mu = \tilde{x}^\mu(x)$ from the Minkowski metric in an inertial coordinate system, $\eta_{\mu\nu}$,
\begin{equation} \label{eq:curvilinearMetricTransformation}
\tilde{g}_{\mu\nu} = \frac{\del x^\rho}{\del \tilde{x}^\mu} \frac{\del x^\sigma}{\del \tilde{x}^\nu} \eta_{\rho\sigma}\,,
\end{equation}
and therefore if one were to evaluate the Riemann curvature tensor using $\tilde{g}_{\mu\nu}$ and $\tilde{\itGamma}^{\lambda}{}_{\mu\nu}$, according to the equation
\begin{equation} \label{eq:RiemannTensor}
R^{\lambda}{}_{\rho\mu\nu} = \del_\mu \tilde{\itGamma}^\lambda{}_{\rho\nu} - \del_\nu \tilde{\itGamma}^\lambda{}_{\rho\mu} + \tilde{\itGamma}^\lambda{}_{\sigma\mu} \tilde{\itGamma}^\sigma{}_{\rho\nu} - \tilde{\itGamma}^\lambda{}_{\sigma\nu} \tilde{\itGamma}^\sigma{}_{\rho\mu}\,,
\end{equation}
one would obtain that $R^{\lambda}{}_{\mu\nu\rho} = 0$ at every point in spacetime, since transforming into a different coordinate system cannot induce the curvature of spacetime.

Now one can apply EP (in this example specifically SEP) in order to couple the scalar field to gravity. The statement of SEP is that, in the presence of a gravitational field (i.e., in curved spacetime), the equation of motion for the scalar field should locally retain the same form as in the absence of the gravitational field (i.e. in flat spacetime). Since equation (\ref{eq:curvilinearKG}) depends only on the field at a given spacetime point and its first and second derivatives at the same point, the equation is in fact local --- it is defined within an infinitesimal neighbourhood of a single point. Given this, EP states that the corresponding equation of motion in the presence of gravity should have precisely the same form:
\begin{equation} \label{eq:curvedKG}
\left( g^{\mu\nu} \nabla_{\mu}\nabla_{\nu} - m^2 \right) \phi(x) = 0\,.	
\end{equation}
The absence of the tilde now denotes the fact that the covariant derivative $\nabla_\mu$ is defined in terms of a generic Levi-Civita connection $\itGamma^\lambda{}_{\mu\nu}$ which is in turn defined in terms of a generic metric $g_{\mu\nu}$, which does not necessarily satisfy (\ref{eq:curvilinearMetricTransformation}). In other words, EP postulates that the equation (\ref{eq:curvedKG}) now holds even in curved spacetime, since for a generic metric and connection, the Riemann curvature tensor need not be equal to zero everywhere. The interaction between the scalar field and gravity, as postulated by EP and implemented in equation (\ref{eq:curvedKG}), is also known in the literature as the {\em minimal coupling} prescription~\cite{bla:02}.

In order to convince oneself that the preparation step of transforming (\ref{eq:flatKG}) to (\ref{eq:curvilinearKG}) is trivial in the sense that it does not introduce any substantial modification of (\ref{eq:flatKG}), one can additionally demonstrate that (\ref{eq:curvedKG}) is in fact locally equivalent to (\ref{eq:flatKG}) as well. Namely, according to a theorem in differential geometry (see for example the end of Chapter 85 in~\cite{lan:lif:80}), at any specific spacetime point $x_0$ one can choose the locally inertial coordinate system, in which the generic metric $g_{\mu\nu}$, the corresponding connection $\itGamma^\lambda{}_{\mu\nu}$ and consequently also the covariant derivative $\nabla_\mu$ take their usual Minkowski values,
\begin{equation} \label{eq:localInertialCoordinates}
g_{\mu\nu}(x_0) = \eta_{\mu\nu}\,, \qquad  \itGamma^\lambda{}_{\mu\nu}(x_0) = 0\,, \qquad \nabla_\mu \Big|_{x=x_0} = \del_\mu\,,
\end{equation}
so that in the infinitesimal neighbourhood of the point $x_0$ equation (\ref{eq:curvedKG}) obtains the form precisely equal to (\ref{eq:flatKG}).

However, note that when {\em integrating} (\ref{eq:curvedKG}), one must take into account that spacetime is curved, since integration is a nonlocal operation, and the locally inertial coordinate system cannot eliminate spacetime curvature. Therefore, the {\em solutions} of (\ref{eq:curvedKG}) will in general be {\em different} from solutions of (\ref{eq:flatKG}), indicating the physical interaction of the scalar field with gravity, despite the fact that the form of the equation of motion is identical in both cases.

Another thing that should be emphasised is that EP itself is not a mathematical theorem, but rather a principle with physical content, since it can be either satisfied of violated. Specifically, we could have prescribed a different coupling of the scalar field to gravity, such that in curved spacetime its equation of motion takes for example the form
\begin{equation} \label{eq:curvedKGviolatedEP}
\left( g^{\mu\nu} \nabla_{\mu}\nabla_{\nu} - m^2 + R^2 + K^2\right) \phi(x) = 0\,,
\end{equation}
where $R \equiv R^{\mu\nu}{}_{\mu\nu}$ and $K \equiv R^{\mu\nu\rho\sigma}R_{\mu\nu\rho\sigma}$ are the curvature scalar and Kretschmann invariant, respectively. This equation is not equivalent to (\ref{eq:curvilinearKG}) and there is no coordinate system in which it can take the form (\ref{eq:flatKG}), since $R$ and $K$ are invariants. In this sense, (\ref{eq:curvedKGviolatedEP}) is an example of a scalar field coupled to gravity such that EP is violated. This type of interaction between matter and gravity is also known in the literature as {\em non-minimal coupling}~\cite{bla:02}.

Finally, we should note that the transformation from (\ref{eq:flatKG}) to (\ref{eq:curvilinearKG}) amounts to what is also known in the literature as {\em symmetry localisation}~\cite{bla:02}. In particular, one can verify that (\ref{eq:flatKG}) remains invariant with respect to the group $\realni^4$ of global translations,
\begin{equation} \label{eq:globalTranslations}
x^\mu \to \tilde{x}^\mu = x^\mu + \zeta^\mu\,, \qquad (\zeta \in \realni^4)\,,
\end{equation}
while (\ref{eq:curvilinearKG}) remains invariant with respect to the group $Dif\!f(\realni^4)$ of spacetime diffeomorphisms, obtained by localisation of the translational symmetry group,
\begin{equation} \label{eq:localTranslations}
x^\mu \to \tilde{x}^\mu = x^\mu + \zeta^\mu(x) \equiv \tilde{x}^\mu(x)\,,
\end{equation}
which represent general curvilinear coordinate transformations, used in (\ref{eq:curvilinearMetricTransformation}). One can explicitly verify that all three equations (\ref{eq:curvilinearKG}), (\ref{eq:curvedKG}) and (\ref{eq:curvedKGviolatedEP}) remain invariant with respect to local translations (\ref{eq:localTranslations}), while describing no coupling to gravity, coupling to gravity that satisfies EP, and coupling to gravity that violates EP, respectively. In this sense, contrary to a common misconception (often stated in the literature) that symmetry localisation gives rise to interactions, one can say that the process of symmetry localisation {\em does not} introduce nor prescribe interactions in any way whatsoever. In particular, a direct counterexample is the equation (\ref{eq:curvilinearMetricTransformation}), which manifestly {\em does} obey local translational symmetry, while it {\em does not} give rise to any interaction whatsoever (see below for the analogous counterexample in electrodynamics).

\subsection{The electromagnetic case}

Let us begin with an example of a Dirac field in Minkowski spacetime, and apply the generalised equivalence principle by coupling it to the EM field. The equation of motion in this case is the ordinary Dirac equation,
\begin{equation} \label{eq:flatD}
\left( i\gamma^{\mu} \del_{\mu} - m \right) \psi(x) = 0\,,
\end{equation}
where $\gamma^{\mu}$ are standard Dirac gamma matrices, satisfying the anticommutator identity of the Clifford algebra $\{\gamma^{\mu},\gamma^{\nu}\} = -2 \eta^{\mu\nu}$. As it stands, equation (\ref{eq:flatD}) describes the free Dirac field, not coupled to an EM field in any way. Note that it is invariant with respect to global $U(1)$ transformations, defined as
\begin{equation} \label{eq:globalUone}
\psi \to \psi' = e^{-i\lambda} \psi \,, \qquad e^{-i\lambda} \in U(1)\,, \qquad \lambda \in\realni\,.
\end{equation}
In order to couple it to standard Maxwell electrodynamics, we first rewrite this equation into a form which is invariant with respect to local $U(1)$ transformations,
\begin{equation} \label{eq:localUone}
\psi \to \psi' = e^{-i\lambda(x)} \psi \,, \qquad \del_\mu \to \tilde{\cD}_\mu = \del_\mu + i \del_\mu \lambda(x)\,,
\end{equation}
so that the equation takes the form
\begin{equation} \label{eq:curvilinearD}
\left( i\gamma^{\mu} \tilde{\cD}_{\mu} - m \right) \psi(x) = 0\,,
\end{equation}
Note that here $\tilde{\cD}$ denotes the covariant derivative with respect to the ``pure gauge'' connection
\begin{equation} \label{eq:UoneTransformation}
\tilde{A}_\mu \equiv \del_\mu \lambda(x) \,,
\end{equation}
where $\lambda(x)$ denotes the arbitrary gauge function. Also note that (\ref{eq:flatD}) is analogous to (\ref{eq:flatKG}), (\ref{eq:curvilinearD}) is analogous to (\ref{eq:curvilinearKG}), while the global and local $U(1)$ gauge transformations (\ref{eq:globalUone}) and (\ref{eq:localUone}) are EM analogs of the global and local spacetime translations (\ref{eq:globalTranslations}) and (\ref{eq:localTranslations}) from the gravitational case. Finally, note that if one were to evaluate the electromagnetic Faraday field strength tensor using $\tilde{A}_\mu$ from (\ref{eq:UoneTransformation}), according to the equation
\begin{equation} \label{eq:FaradayTensor}
F_{\mu\nu} = \del_\mu \tilde{A}_\nu - \del_\nu \tilde{A}_\mu \,,
\end{equation}
one would obtain that $F_{\mu\nu} = 0$ at every point in spacetime, since the potential that is a pure gauge cannot induce an EM field. Here (\ref{eq:FaradayTensor}) is analogous to (\ref{eq:RiemannTensor}).

Once the Dirac equation in the form (\ref{eq:curvilinearD}) is in hand, one can apply the electromagnetic generalisation of EP in order to couple the Dirac field to an EM field. The statement of EP in this case is that, in the presence of an EM field, the equation of motion for the Dirac field should locally retain the same form as in the absence of the EM field. Since equation (\ref{eq:curvilinearD}) depends only on the field at a given spacetime point and its first derivatives at the same point, it is therefore defined within an infinitesimal neighbourhood of a single point --- in other words, it is local. Given this, electromagnetic EP states that the corresponding equation of motion in the presence of EM field should have precisely the same form (the analog of (\ref{eq:curvedKG})):
\begin{equation} \label{eq:curvedD}
\left( i\gamma^{\mu} \cD_{\mu} - m \right) \psi(x) = 0\,.
\end{equation}
The absence of the tilde now denotes the fact that the covariant derivative $\cD_\mu \equiv \del_\mu + i A_\mu $ is defined in terms of a generic $U(1)$ connection $A_\mu$ which does not necessarily satisfy (\ref{eq:UoneTransformation}), but does obey the usual gauge transformation rule,
\begin{equation} \label{eq:connectionUoneTransformation}
A_\mu \to A'_\mu = A_\mu + \del_\mu \lambda(x)\,.
\end{equation}
In other words, electromagnetic EP postulates that the equation (\ref{eq:curvedD}) holds even in the presence of an EM field, since for a generic connection $A_\mu$ the Faraday tensor may not be equal to zero everywhere.  The interaction between the Dirac field and the EM field as postulated by the electromagnetic EP and implemented in equation (\ref{eq:curvedD}) is again known in the literature as the {\em minimal coupling} prescription~\cite{pes:sch:95,bla:02}.

If one wishes to convince oneself that the preparation step of transforming (\ref{eq:flatD}) to (\ref{eq:curvilinearD}) is trivial in the sense that it does not introduce any substantial modification of (\ref{eq:flatD}), one can additionally demonstrate that (\ref{eq:curvedD}) is in fact locally equivalent to (\ref{eq:flatD}). To do this, at any specific spacetime point $x_0$ one can choose the following $U(1)$ gauge,
\begin{equation} \label{eq:localUoneGauge}
\lambda(x) = - A_\mu(x_0) x^\mu\,,
\end{equation}
so that, according to (\ref{eq:connectionUoneTransformation})
\begin{equation}
A'_\mu (x) = A_\mu(x) - \del_\mu \left( A_\nu(x_0) x^\nu \right) \qquad \Rightarrow \qquad A'_\mu(x_0) = 0\,, \qquad \cD_\mu \Big|_{x=x_0} = \del_\mu\,.
\end{equation}
This choice of gauge is the EM analog of the choice of a locally inertial coordinate system (\ref{eq:localInertialCoordinates}). Substituting this into the primed version of (\ref{eq:curvedD}) and evaluating the whole equation at $x=x_0$, it reduces precisely to the form (\ref{eq:flatD}) in the infinitesimal neighbourhood at that point, despite the presence of nonzero EM field.

Again note that when {\em integrating} (\ref{eq:curvedD}), one must take into account that EM field is nonzero, since integration is a nonlocal operation, and the choice of gauge (\ref{eq:localUoneGauge}) eliminates the EM potential from (\ref{eq:curvedD}) only at the point $x_0$, while the Faraday tensor is gauge invariant. Therefore, the {\em solutions} of (\ref{eq:curvedD}) will in general be {\em different} from solutions of (\ref{eq:flatD}), indicating the physical interaction of the Dirac field with EM field, despite the fact that the form of the equation of motion for the Dirac field is identical in both cases.

As in the case of gravity, we should emphasise that the electromagnetic EP is not a mathematical theorem, but rather a principle with physical content, since it can be either satisfied of violated. Specifically, we could have prescribed a different coupling of the Dirac field to electrodynamics, such that in the presence of an EM field its equation of motion takes for example the form (analog of (\ref{eq:curvedKGviolatedEP}))
\begin{equation} \label{eq:curvedDviolatedEP}
\left( i\gamma^{\mu} \cD_{\mu} - m + I_1 + I_2 \right) \psi(x) = 0\,,
\end{equation}
where $I_1 \equiv F^{\mu\nu}F_{\mu\nu}$ and $I_2 \equiv \lc^{\mu\nu\rho\sigma}F_{\mu\nu}F_{\rho\sigma}$ are the two fundamental invariants of the Faraday tensor. This equation is not equivalent to (\ref{eq:curvilinearD}), and there exists no local $U(1)$ gauge in which it could take the form (\ref{eq:flatD}), since $I_1$ and $I_2$ are invariants. In this sense, (\ref{eq:curvedDviolatedEP}) is an example of a Dirac field coupled to EM field such that the electromagnetic EP is violated. This is also known in the literature as {\em non-minimal coupling}~\cite{pes:sch:95,bla:02}.

Finally, we should also note that the transformation from (\ref{eq:flatD}) to (\ref{eq:curvilinearD}) amounts to what is also known in the literature as {\em symmetry localisation}~\cite{pes:sch:95,bla:02}. Specifically, one can explicitly verify that all three equations (\ref{eq:curvilinearD}), (\ref{eq:curvedD}) and (\ref{eq:curvedDviolatedEP}) remain invariant with respect to local $U(1)$ gauge transformations, while describing no coupling to an EM field, coupling to an EM field that satisfies the electromagnetic EP, and coupling to an EM field that violates electromagnetic EP, respectively. In this sense, one can again say that the process of symmetry localisation {\em does not} introduce nor prescribe interactions in any way whatsoever. In the case of electrodynamics and other gauge theories, this is quite often misrepresented in literature --- the step of symmetry localisation is silently joined together with the step of applying the electromagnetic version of EP, thus in the end giving rise to an interacting theory, and the resulting presence of the interaction is then mistakenly attributed to the localisation of symmetry, rather than to the application of EP. Similar to the gravitational case above, the equation of motion (\ref{eq:curvilinearD}) is an explicit counterexample to such an attribution, since it {\em does} have local $U(1)$ symmetry, but {\em does not} have any interaction with an EM field.

\subsection{The test particle case}

The last topic we should address is the context in which the statement of electromagnetic EP is compatible with the existence of the Lorentz force law, acting on charged test particles. Namely, one often distinguishes the motion of a test particle in a gravitational field from a motion of a test particle in an EM field, by comparing the geodesic equation (\ref{eq:geodesic_kink})
\begin{equation} \label{eq:geodesic_kink_App}
u^\mu(\tau) \nabla_\mu u^\lambda(\tau) = 0\,,
\end{equation}
where $u^{\mu}$ is the $4$-velocity of the test particle, with the Lorentz force equation
\begin{equation} \label{eq:LorentzForce}
u^\mu(\tau) \nabla_\mu u^\lambda(\tau) = \frac{q}{m} F^{\lambda\rho}\,u_{\rho}(\tau)\,,
\end{equation}
where $q/m$ is the charge to mass ratio of a test particle moving in an external EM field, described by the Faraday tensor $F_{\mu\nu}$. A typical conclusion one draws from this comparison is that the interaction with the EM field gives rise to a ``real force'', while the interaction with the gravitational field does not.

However, it is highly misleading to compare (\ref{eq:geodesic_kink_App}) to (\ref{eq:LorentzForce}) in the first place. Namely, as we have discussed in detail in Section \ref{sec:particle_in_gr}, in field theory the notion of a particle can be defined only approximately, and this applies equally for electrodynamics as well as for gravity. Specifically, given the example discussed above, of a Dirac field coupled to an EM field via equation (\ref{eq:curvedD}), we have seen that in the infinitesimal neighbourhood of a given point $x_0$ one can completely gauge away any presence of the coupling to EM field from (\ref{eq:curvedD}). In this sense, the notion of a test particle that satisfies (\ref{eq:LorentzForce}) cannot be identified with an idealised point-particle, that has exactly zero size. Instead, the realistic test particle is a wave-packet configuration of a Dirac field (a kink), and as such has small but nonzero size. As it evolves, the different parts of the wave-packet are subject to interaction with the EM potential $A_\mu$ at {\em different} points of spacetime, giving rise to an effective non-minimal coupling with the Faraday tensor $F_{\mu\nu}$. This is completely analogous to the case of a test particle with small but nonzero size interacting with spacetime curvature due to tidal forces --- both effects are equally nonlocal, since both kinks have nonzero size. On the other hand, a test particle that satisfies (\ref{eq:geodesic_kink_App}) represents an idealised point-particle (a leading order approximation in the multipole expansion of the matter field), i.e., a kink which thus has precisely zero size.

In this sense, the Lorentz force equation (\ref{eq:LorentzForce}) rather ought to be compared with the Papapetrou equation (\ref{eq:PapapetrouEq}),
\begin{equation} \label{eq:PapapetrouEqApp}
u^\mu(\tau) \nabla_\mu u^\lambda(\tau) = R^\lambda{}_{\mu\rho\sigma}\, u^\mu(\tau) J^{\rho\sigma}(\tau) \,.
\end{equation}
And indeed, one can see quite a reasonable analogy between (\ref{eq:LorentzForce}) and (\ref{eq:PapapetrouEqApp}). There are of course small technical differences due to the precise nature of the coupling to various moments of the kink, but nevertheless, the two equations are strikingly similar. Given this, while one can still draw the conclusion that the interaction of a kink with the EM field gives rise to a ``real force'', one can draw precisely the same conclusion for the interaction of a kink with the gravitational field. There is no distinction between gravity and the other gauge interactions at this level --- all four interactions in nature (strong, weak, electromagnetic and gravitational) are equally ``real''. In addition, all four interactions satisfy EP at the fundamental field theory level (i.e., in the sense of strong generalised EP), while at the level of mechanics, a corresponding weak generalised EP is manifestly violated in all four cases.
}

\bibliography{wep-violation-paper}

\begin{thebibliography}{10}
\def\enquote#1{``#1''}
\expandafter\ifx\csname url\endcsname\relax
  \def\url#1{{\tt #1}}\fi
\expandafter\ifx\csname urlprefix\endcsname\relax\def\urlprefix{URL }\fi
\expandafter\ifx\csname eprint\endcsname\relax\def\eprint#1{\url{#1}}\fi

\bibitem{kay:98}
B.~S. Kay, {\it Class. Quant. Grav.\/} {\bf 15}, L89 (1998),
  \eprint{arXiv:hep-th/9810077}.

\bibitem{oni:wan:16}
T.~Oniga and C.~H.~T. Wang, {\it Phys. Rev. D\/} {\bf 93}, 044027 (2016),
  \eprint{arXiv:1511.06678}.

\bibitem{bru:16}
D.~E. Bruschi, {\it Phys. Lett. B\/} {\bf 754}, 182 (2016),
  \eprint{arXiv:1412.4007}.

\bibitem{bos:etal:17}
S.~Bose, A.~Mazumdar, G.~W. Morley, H.~Ulbricht, M.~Toro{\v{s}},
  M.~Paternostro, A.~A. Geraci, P.~F. Barker, M.~S. Kim and G.~Milburn, {\it
  Phys. Rev. Lett.\/} {\bf 119}, 240401 (2017), \eprint{arXiv:1707.06050}.

\bibitem{mar:ved:17}
C.~Marletto and V.~Vedral, {\it Phys. Rev. Lett.\/} {\bf 119}, 240402 (2017),
  \eprint{arXiv:1707.06036}.

\bibitem{mar:ved:18}
C.~Marletto and V.~Vedral, {\it Phys. Rev. D\/} {\bf 98}, 046001 (2018).

\bibitem{pau:voj:18}
N.~Paunkovi\'c and M.~Vojinovi\'c, {\it Class. Quant. Grav.\/} {\bf 35}, 185015
  (2018), \eprint{arXiv:1702.07744}.

\bibitem{ore:cos:bru:12}
O.~Oreshkov, F.~Costa and {\v{C}}.~Brukner, {\it Nature Communications\/} {\bf
  3}, 1092 (2012).

\bibitem{ara:bra:cos:fei:gia:bru:15}
M.~Ara{\'u}jo, C.~Branciard, F.~Costa, A.~Feix, C.~Giarmatzi and {\v
  C}.~Brukner, {\it New Journal of Physics\/} {\bf 17}, 102001 (2015).

\bibitem{vil:17}
V.~Vilasini, {\it An introduction to causality in quantum theory (and
  beyond)\/}, Master's thesis, ETH, Z\"urich, Switzerland (2017).

\bibitem{ore:19}
O.~Oreshkov, {\it Quantum\/} {\bf 3}, 206 (2019).

\bibitem{pau:voj:20}
N.~Paunkovi{\'c} and M.~Vojinovi{\'c}, {\it Quantum\/} {\bf 4}, 275 (2020),
  \eprint{arXiv:1905.09682}.

\bibitem{vil:col:22}
V.~Vilasini and R.~Colbeck, {\it Phys. Rev. A\/} {\bf 106}, 032204 (2022),
  \eprint{arXiv:2109.12128}.

\bibitem{vil:ren:22}
V.~Vilasini and R.~Renner, {\it {\tt arXiv:2203.11245}\/} .

\bibitem{orm:van:bar:22}
N.~Ormrod, A.~Vanrietvelde and J.~Barrett, {\it {\tt arXiv:2204.10273}\/} .

\bibitem{gia:cas:bru:19}
F.~Giacomini, E.~Castro-Ruiz and {\v{C}}.~Brukner, {\it Nat. Comm.\/} {\bf 10},
  1 (2019).

\bibitem{van:hoh:gia:cas:20}
A.~Vanrietvelde, P.~A. H{\"o}hn, F.~Giacomini and E.~Castro-Ruiz, {\it
  Quantum\/} {\bf 4}, 225 (2020), \eprint{arXiv:1809.00556}.

\bibitem{kru:hoh:mul:21}
M.~Krumm, P.~A. H{\"o}hn and M.~P. M{\"u}ller, {\it Quantum\/} {\bf 5}, 530
  (2021), \eprint{arXiv:2011.01951}.

\bibitem{ahm:etal:22}
S.~A. Ahmad, T.~D. Galley, P.~A. H\"ohn, M.~P.~E. Lock and A.~R.~H. Smith, {\it
  Phys. Rev. Lett.\/} {\bf 128}, 170401 (2022), \eprint{arXiv:2103.01232}.

\bibitem{ham:kab:cas:bru:21}
A.-C. de~la Hamette, V.~Kabel, E.~Castro-Ruiz and {\v{C}}.~Brukner, {\it {\tt
  arXiv:2112.11473}\/} .

\bibitem{col:kos:98}
D.~Colladay and V.~A. Kosteleck\'y, {\it Phys. Rev. D\/} {\bf 58}, 116002
  (1998), \eprint{hep-ph/9809521}.

\bibitem{kos:rus:11}
V.~A. Kosteleck\'y and N.~Russell, {\it Rev. Mod. Phys.\/} {\bf 83}, 11 (2011),
  \eprint{arXiv:0801.0287}.

\bibitem{ame:12}
G.~Amelino-Camelia, {\it Symmetry\/} {\bf 4}, 344 (2012),
  \eprint{arXiv:1111.5643}.

\bibitem{ame:pal:ron:ami:20}
G.~Amelino-Camelia, M.~Palmisano, M.~Ronco and G.~D’Amico, {\it Int. Jour.
  Mod. Phys. D\/} {\bf 29}, 2050017 (2020), \eprint{arXiv:1910.05997}.

\bibitem{tor:ant:mir:19}
M.~D.~C. Torri, V.~Antonelli and L.~Miramonti, {\it Eur. Phys. Jour. C\/} {\bf
  79}, 1 (2019), \eprint{arXiv:1906.05595}.

\bibitem{pip:pau:voj:19}
F.~Pipa, N.~Paunkovi{\'c} and M.~Vojinovi{\'c}, {\it Journal of Cosmology and
  Astroparticle Physics\/} {\bf 2019}, 057 (2019), \eprint{arXiv:1801.03207}.

\bibitem{gia:bru:20}
F.~Giacomini and {\v{C}}.~Brukner, {\it {\tt arXiv:2012.13754}\/} .

\bibitem{gia:bru:22}
F.~Giacomini and {\v{C}}.~Brukner, {\it AVS Quantum Science\/} {\bf 4}, 015601
  (2022), \eprint{arXiv:2109.01405}.

\bibitem{mar:ved:20}
C.~Marletto and V.~Vedral, {\it Frontiers in Physics\/} {\bf 8}, 176 (2020),
  \eprint{arXiv:2004.11616}.

\bibitem{mar:ved:21}
C.~Marletto and V.~Vedral, {\it Jour. Phys. Commun.\/} {\bf 5}, 051001 (2021),
  \eprint{arXiv:2001.02777}.

\bibitem{mar:ved:22}
C.~Marletto and V.~Vedral, {\it AVS Quantum Science\/} {\bf 4}, 015603 (2022),
  \eprint{arXiv:2005.00138}.

\bibitem{EinsteinInfeldHoffmann}
A.~Einstein, L.~Infeld and B.~Hoffmann, {\it Ann. Math.\/} {\bf 39}, 65 (1938).

\bibitem{Mathisson}
M.~Mathisson, {\it Acta Phys. Pol.\/} {\bf 6}, 163 (1937).

\bibitem{Papapetrou}
A.~Papapetrou, {\it Proc. R. Soc. A\/} {\bf 209}, 248 (1951).

\bibitem{Tulczyjew}
W.~Tulczyjev, {\it Acta Phys. Pol.\/} {\bf 18}, 393 (1959).

\bibitem{Taub}
A.~H. Taub, {\it Jour. Math. Phys.\/} {\bf 5}, 112 (1964).

\bibitem{Dixon1}
G.~Dixon, {\it Nuovo Cim.\/} {\bf 34}, 317 (1964).

\bibitem{Dixon2}
G.~Dixon, {\it Nuovo Cim.\/} {\bf 38}, 1616 (1965).

\bibitem{Dixon3}
G.~Dixon, {\it Proc. R. Soc. A\/} {\bf 314}, 499 (1970).

\bibitem{Dixon4}
G.~Dixon, {\it Proc. R. Soc. A\/} {\bf 319}, 509 (1970).

\bibitem{Dixon5}
G.~Dixon, {\it Gen. Relativ. Gravit.\/} {\bf 4}, 199 (1973).

\bibitem{YasskinStoeger}
P.~B. Yasskin and W.~R. Stoeger, {\it Phys. Rev. D\/} {\bf 21}, 2081 (1980).

\bibitem{ShirafujiNomuraHayashiOne}
K.~Nomura, T.~Shirafuji and K.~Hayashi, {\it Prog. Theor. Phys.\/} {\bf 86},
  1239 (1991).

\bibitem{ShirafujiNomuraHayashiTwo}
K.~Nomura, T.~Shirafuji and K.~Hayashi, {\it Prog. Theor. Phys.\/} {\bf 87},
  1275 (1992).

\bibitem{VasilicVojinovicJHEP}
M.~Vasili\'c and M.~Vojinovi\'c, {\it JHEP\/} {\bf 07}, 028 (2007),
  \eprint{arXiv:0707.3395}.

\bibitem{VasilicVojinovicCetiri}
M.~Vasili\'c and M.~Vojinovi\'c, {\it Phys. Rev. D\/} {\bf 78}, 104002 (2008),
  \eprint{arXiv:1010.1861}.

\bibitem{AcciolyPaszko}
A.~Accioly and R.~Paszko, {\it Adv. Studies Theor. Phys.\/} {\bf 3}, 65 (2009).

\bibitem{Longhi}
S.~Longhi, {\it Opt. Lett.\/} {\bf 43}, 226 (2018), \eprint{arXiv:1712.02054}.

\bibitem{Chowdhury}
P.~Chowdhury, D.~Home, A.~S. Majumdar, S.~V. Mousavi, M.~R. Mozaffari and
  S.~Sinha, {\it Class. Quant. Grav.\/} {\bf 29}, 025010 (2012),
  \eprint{arXiv:1107.5159}.

\bibitem{Rosi}
G.~Rosi, G.~D’Amico, L.~Cacciapuoti, F.~Sorrentino, M.~Prevedelli, M.~Zych,
  {\v C}.~Brukner and G.~M. Tino, {\it Nat. Comm.\/} {\bf 8}, 15529 (2017),
  \eprint{arXiv:1704.02296}.

\bibitem{zyc:bru:17}
M.~Zych and {\v C}.~Brukner, {\it Nature Phys.\/} {\bf 14}, 1027 (2018),
  \eprint{arXiv:1502.00971}.

\bibitem{ana:hu:18}
C.~Anastopoulos and B.~L. Hu, {\it Class. Quant. Grav.\/} {\bf 35}, 035011
  (2018), \eprint{arXiv:1707.04526}.

\bibitem{har:20}
L.~Hardy, in D.~Finster, F.and~Giulini, J.~Kleiner and J.~Tolksdorf, eds.,
  \enquote{Progress and Visions in Quantum Theory in View of Gravity},
  189--220, Springer International Publishing, Cham (2020),
  \eprint{arXiv:1903.01289}.

\bibitem{kretschmann}
E.~Kretschmann, {\it Annalen der Physik\/} {\bf 358}, 575 (1918).

\bibitem{mis:tho:whe:73}
C.~W. Misner, K.~S. Thorne and J.~A. Wheeler, {\it Gravitation\/}, W. H.
  Freeman and Co., San Francisco (1973).

\bibitem{oco:cal:11}
E.~Okon and C.~Callender, {\it Eur. Jour. Phil. Sci.\/} {\bf 1}, 133 (2011),
  \eprint{arXiv:1008.5192}.

\bibitem{cas:lib:son:15}
E.~D. Casola, S.~Liberati and S.~Sonego, {\it Am. J. Phys.\/} {\bf 83}, 39
  (2015), \eprint{arXiv:1310.7426}.

\bibitem{vio:ono:97}
L.~Viola and R.~Onofrio, {\it Phys. Rev. D\/} {\bf 55}, 455 (1997),
  \eprint{arXiv:quant-ph/9612039}.

\bibitem{ply:98}
R.~Plyatsko, {\it Phys. Rev. D\/} {\bf 58}, 084031 (1998).

\bibitem{bos:maz:sch:tor:22}
S.~Bose, A.~Mazumdar, M.~Schut and M.~Toro{\v{s}}, {\it {\tt
  arXiv:2203.11628}\/} .

\bibitem{bla:02}
M.~Blagojevi\'c, {\it {Gravitation and Gauge Symmetries}\/}, Institute of
  Physics Publishing, Bristol (2002).

\bibitem{lan:lif:80}
L.~D. Landau and E.~M. Lifshitz, {\it {The Classical Theory of Fields (4th
  edition)}\/}, Butterworth-Heinemann, Oxford (1980).

\bibitem{pes:sch:95}
M.~E. Peskin and D.~V. Schroeder, {\it {An Introduction to Quantum Field
  Theory}\/}, Addison-Wesley Publishing Co., Boston (1995).

\end{thebibliography}

\end{document}